\documentstyle[12pt]{article}
\newcommand{\mysection}[1]{\setcounter{equation}{0}\section{#1}}

\begin{document}
\setlength{\baselineskip}{0.30in}

\newcommand{\beq}{\begin{equation}}
\newcommand{\eeq}{\end{equation}}
\newcommand{\beqa}{\begin{eqnarray}}
\newcommand{\eeqa}{\end{eqnarray}}
\newcommand{\ls}{L\"{u}scher-Schechter\,\,}
\newcommand{\lsim}{\begin{array}{c}\,\sim\vspace{-21pt}\\<\end{array}}
\newcommand{\gsim}{\begin{array}{c}\sim\vspace{-21pt}\\>\end{array}}
\newcommand{\bfk}{{\bf k}}
\newcommand{\ak} {a(\bfk )}
\newcommand{\aks}{a^*(\bfk )}
\newcommand{\bk} {b(\bfk )}
\newcommand{\bks}{b^*(\bfk )}
\newcommand{\once}{{8\pi^2\over g^2}}
\newcommand{\twice}{{16\pi^2\over g^2}}

\begin{titlepage}
{\hbox to\hsize{April 1993 \hfill {\hspace{1.5cm} JHU-TIPAC-930012} \hfill
hep-ph/9305263 }}
\begin{center}
\vglue .06in
{\Large \bf Semiclassical Scattering in Yang-Mills Theory}
     \\[.5in]

\begin{tabular}{c}
{\bf Thomas M. Gould\footnotemark[1]\footnotemark[2]\footnotemark[3]}
and {\bf Erich R. Poppitz\footnotemark[1]\footnotemark[4]}\\[.05in]
{\it Department of Physics and Astronomy}\\
{\it The Johns Hopkins University}\\
{\it Baltimore MD 21218 }\\[.15in]
\end{tabular}

{Abstract}\\[-.05in]

\footnotetext[1]{Supported in part by the National Science Foundation
under grant PHY-90-9619.}
\footnotetext[2]{Supported by the Texas National Research Laboratory
Commission under grant RGFY-93-292.}
\footnotetext[3]{gould@fermi.pha.jhu.edu}
\footnotetext[4]{poppitz@dirac.pha.jhu.edu}
$\bigtriangledown $
\end{center}

\begin{quotation}
A classical solution to the Yang-Mills theory is given a new semiclassical
interpretation.
The boundary value problem on a complex time contour
which arises from the semiclassical approximation to multiparticle scattering
amplitudes is reviewed and applied to the case of Yang-Mills theory.
The solution describes a classically forbidden transition between
states with a large average number of particles in the limit $g\rightarrow 0$.
It dominates a transition probability with a semiclassical
suppression factor equal to twice the action of the well-known BPST instanton.
Hence, it is relevant to the problem of high energy tunnelling.
It describes transitions of unit topological charge for an
appropriate time contour.
Therefore, it may have a direct interpretation in terms of fermion number
violating processes in electroweak theory.
The solution describes a transition between an initial state with
parametrically fewer particles than the final state.
Thus, it may be relevant to the study of semiclassical initial
state corrections in the limit of a small number of initial particles.
The implications of these results for multiparticle production in
electroweak theory are also discussed.
\end{quotation}
\end{titlepage}
\newpage

\section{Introduction}

An intriguing feature of the Yang-Mills gauge theory is the periodic
structure of its vacuum \cite{JAC,CDG}.
In the semiclassical approximation,
the topology of finite energy solutions leads to a classification
of all gauge-inequivalent vacua in the theory.
The discovery of this rich structure has had a profound impact on our
understanding of non-perturbative aspects of the theory,
notably  low energy phenomena like the solution of the famous U(1)
problem in QCD \cite{THO}.
However,
the role of the vacuum in the dynamics of particle scattering,
and in particular high energy multiparticle scattering,
is not yet  as deeply understood.
This deficit in our understanding has been confronted in recent years
 with the study of so-called ``instanton-induced'' cross-sections
\cite{RIN,ESP,early,MAT}.

The simplest semiclassical estimate of the contribution of
the BPST  instanton \cite{BPST} to a total inclusive two particle cross-section
in electroweak theory implies a result which grows exponentially with
center-of-mass energy \cite{RIN,ESP,MAT,KRT}.
The same behavior has also appeared in a large number of model field
theories with instanton solutions throughout an extensive series of
investigations \cite{early,MAT}.
It has further been shown that the instanton is the basis of
a systematic perturbative expansion of the final state radiative
corrections to the cross-section \cite{KRT}.
This determines the leading semiclassical behavior,
neglecting initial state radiative corrections,
\beq
\label{eq:holygrail}
\sigma_{tot} (x) \; \sim \;
\exp{\left[\, \twice F(x) \, + \, o(\alpha^0)\,\right]}
\eeq
as an expansion in powers of a small parameter, $x\equiv E/E_0$,
the ratio of the center-of-mass energy $E$ and a mass scale of
order the electroweak sphaleron mass,
$E_0 \simeq M_w/\alpha_w \simeq 10\;{\rm TeV}$.
The so-called ``Holy Grail function'', $F(x)$, is approximately
$-1$ for small $x$, reflecting the severe\, 'tHooft suppression factor,
$\exp{-\twice} \simeq 10^{-127}$, due to the large instanton action.
The fact that the Holy Grail function is an increasing function of $x$
for small $x$ has led many to speculate about the possibility of overcoming
the severe exponential suppression factor at energies of order $E_0 $.
The possibility of strong multiparticle scattering in electroweak theory
at  energies in the multi-TeV range has led to an enormous effort to understand
the behavior of multiparticle cross-sections in the sphaleron
energy regime \cite{MAT}.

One approach has been to consider a mechanism by which the exponentially
growing cross-section unitarizes at high energies.
In this regard,
multi-instanton contributions to the cross-section (\ref{eq:holygrail})
have been considered,
as a means to unitarize the cross-section by $s$-channel
iteration of the one-instanton contribution \cite{ZAK,MAG,GOU}.
It has been argued that if one assumes the validity of (\ref{eq:holygrail})
in the one-instanton sector,
then strong multi-instanton contributions become important before
the\, 'tHooft suppression is overcome.

However,
all of these conclusions are based on semiclassical expansions around
configurations which are not influenced by external sources.
Indeed,
the instanton and multi-instantons obey vacuum boundary conditions,
and as such are relevant to this problem only in the approximation in
which external sources are neglected.
While the final state corrections can be taken into account in
the perturbative
expansion in $x$,
the initial state corrections are more subtle\footnote{In addition,
the distinction between corrections involving initial and final state particles
is ambiguous at high orders of the low energy expansion \cite{KT}.}.
These involve radiative corrections to hard particles which are not
{\it a priori} expressible semiclassically.
However,
there have been some indications \cite{MUE,MMY,TIN} that the contributions to
$F(x)$ of corrections involving hard initial legs may also be calculable in a
semiclassical manner.
It may then be possible to calculate the entire leading order semiclassical
exponent
in a saddle point approximation.
What is needed  is a new technique which accounts for external sources
to make the semiclassical behavior of the total cross-section manifest.

A strategy for out-flanking the problem of initial state corrections
was recently proposed by Rubakov, Son and Tinyakov \cite{TIN,RST,RT,RSTnew}.
The basic idea is to consider transitions from states of a fixed large
number of particles, say $N_{\rm in} = \nu/g^2$.
The instanton-induced transition probability from a multiparticle initial state
is
then calculable semiclassically, in the limit $g \rightarrow 0$ with
$\nu$ fixed.
Its leading semiclassical behavior is  determined by the solution to a boundary
value problem.
The boundary conditions imposed at initial and final times correctly account
for
the energy transfer from the initial multiparticle state to the final
multiparticle state.
The leading semiclassical behavior of the $N_{\rm in}$-particle transition
probability has
a form similar to (\ref{eq:holygrail})
\beq
\label{eq:nparticle}
\sigma_{N_{\rm in}} (x) \; \sim \;
\exp{\left[\, -\twice F(x,\nu) \, + \, o(g^0)\,\right]}  \, .
\eeq
The function $F(x, \nu)$ is a rigorous upper bound on the two-particle ``Holy
Grail''
function (\ref{eq:holygrail}),
\beq
\label{bound}
F(x, \nu) \; \gsim \; F(x)  \hspace{2cm} x \equiv E/E_0 \, ,
\eeq
and is related to a lower bound under less rigorous assumptions \cite{TIN}.
Since this function contains all initial state corrections for the
$N_{\rm in} = \nu/g^2$ particle transition,
it is hoped that it reproduces the leading semiclassical behavior
of two-particle transition when $\nu$ is small,
including initial and final state corrections.
Initial indications from explicit calculations of initial and final
state corrections are that the limit  $\nu\rightarrow 0$ is smooth
\cite{MUE,TIN},
so that the contribution to a semiclassical transition probability
from the solution of the boundary value problem contains the initial
and final state corrections.
The boundary value problem posed in this way also holds the promise of
being amenable in principle to numerical computation of multiparticle
transitions.
It would now be useful to have some analytical examples to guide future
efforts in this direction \cite{RSTnew,KYA}.

For the calculation of an instanton-induced (i.e. tunneling) transition at
fixed energy,
the choice of a Minkowski or Euclidean time contour is too restrictive.
The boundary value problem is instead conveniently formulated on
a {\it complex} time contour, to be explained below.
A few such solutions on a complex time contour have already been investigated.

A classical solution with two turning points on a complex time contour is the
so-called
{\it periodic instanton} \cite{KRT3}
\footnote{This should not be confused with the periodic solution
to the Euclidean formulation of finite-temperature Yang-Mills theory
\cite{GPY},
nor with the periodic multi-instanton configurations \cite{ZAK,MAG,GOU}.}.
The periodic instanton is a solution to the complex-time boundary value problem
which arises in the semiclassical approximation to the inclusive transition
probability from all initial states at fixed energy,
or a microcanonical distribution.
It has been shown to determine the maximal probability for transition
in the one-instanton sector from states of fixed energy \cite{KRT3,RT}.
The periodic instanton in electroweak theory has so far been
constructed only in a low energy approximation,
and the resulting transition probability is determined in
a perturbative expansion similar to that in (\ref{eq:holygrail}).
It has been found to describe transitions between states of equal number of
particles
which is large in the semiclassical limit, $N_{\rm in} = N_{\rm
fin} \sim 1/g^2$.
So, this solution is irrelevant for describing $2 \rightarrow n$ scattering
processes at high energies,
though it does play a role in determining the rate of tunnelling,
and anomalous baryon number violation, at finite temperature \cite{HSU}.

A solution which describes transitions from a state of smaller number of
particles to a state with a larger number of particles has also  been
constructed in a low energy
expansion \cite{RT}.
Similarly,
it determines the maximum transition probability from states of
fixed energy and particle number.
However,
it remains to construct solutions which describe such processes in general.
This is a formidable task, requiring a solution of the Yang-Mills equations
with arbitrary boundary conditions on a complex time contour.
In this paper, we pursue more modest goals.
We investigate the properties of a well-known,
highly symmetric Minkowski time solution on a complex time contour.
The solution in Minkowski time describes an energy density which
evolves from early times as a thin collapsing spherical shell,
bounces at an intermediate time,
and expands outward again at late times.
As yet,
the role of this solution in scattering problems has not been
fully developed \cite{FKS}.

We show that a subclass of the SO(4)-conformally invariant
solutions found by L\"uscher and Schechter exhibits a number of remarkable
properties on a suitably chosen complex time contour:

\begin{enumerate}
\item The semiclassical suppression is equal to the action of the
BPST instanton, \mbox{${\rm Im} \,  S = \once$.}
This quantity controls the semiclassical exponential
dependence of a transition probability between coherent states.
\item The topological charge of the solution is equal to
the BPST instanton charge, \mbox{$Q = 1$.}
Thus,
the solution may have a direct interpretation for
fermion number violating processes \cite{FKS}.
\item It solves the boundary value problem for the transition
probability in the one-instanton sector from a coherent state
with a smaller number of particles, to a state with a larger number
of particles. This property makes the solution interesting for
the investigation of $2\rightarrow n$ processes with fermion number violation
at high energies\cite{TIN,RST,RT}.
\end{enumerate}

Thus,
the L\"{u}scher-Schechter solution considered in this paper provides
an analytical benchmark for future numerical computations of
many particle transition amplitudes in Yang-Mills theory.
The work presented here makes explicit use of the conformal invariance
of the Yang-Mills theory,
both in the construction of the solution and in the calculation of its
properties.
To the extent that this theory represents the high energy limit of
a spontaneously-broken gauge theory,
the results also have implications for multiparticle cross-sections and
high energy baryon number violation in electroweak theory.
These implications will be discussed in the final section of this paper.

This paper is organized as follows.
In Section 2,
we show that the imaginary part of the action and
the topological charge of the solution are determined
solely by the number $N$ of singularities of the solution
in the complex time plane, enclosed between the complex time
contour and the real time axis.
These solutions have therefore the remarkable
property that the imaginary part of the action and the
topological charge obey
$$
{g^2 \over 8\pi^2}\, {\rm Im}\, S \: = \: Q \: = \: N .
$$
In Section 3,
we show that the initial and final coherent states
contain a different number of particles,
the ratio being controlled by a parameter of the solution.
We also explicitly demonstrate that the initial and final gauge field
configurations
belong to different topological sectors,
in agreement with the results of Section 2.
In Section 4,
we discuss the relationship of this classical solution to the saddle points
of transition probabilities at fixed energy and particle number.
Finally,
Section 5 contains a discussion of the results and an outlook on unsolved
problems.

\mysection{The Action and Topological Charge of the Solution}

In this section,
we describe the classical solution and an appropriate
choice of the complex time contour.
Then, we compute its action and topological charge.
The use of a Minkowski or Euclidean time contour for the semiclassical
calculation
of transition amplitudes in the one-instanton sector is too restrictive.
Recall that computing tunneling contributions  to fixed-energy (i.e.
time-independent)
Green functions in quantum mechanics can be performed in the WKB approximation
only on
a complex time contour,
chosen to lie in Minkowski directions at early and late times,
with a period of Euclidean evolution inserted at an intermediate time.
They give the dominant WKB-contribution to classically forbidden processes.
In the present case of quantum field theory,
we will similarly be interested only in time-independent transition
probabilities.

In the case of nonabelian gauge theories,
the transitions between vacua with different topological number,
signalled by fermion number or chirality violation,
are analogous to the classically forbidden processes in quantum mechanics.
The vacuum-to-vacuum transition amplitude is known to be maximized by
instantons,
in the semiclassical approximation.
They, and in fact any finite action Euclidean solution,
satisfy vacuum boundary conditions at infinity.
However,
for transitions involving many-particle initial and final states,
vacuum boundary conditions are clearly not the correct ones.
Considering solutions on a complex time contour, $C_T$, (fig.1) provides
a natural description of the initial and final states in Minkowski space in
terms of
the free wave asymptotics of the solution at
$\vert\, {\rm Re}\, t\,\vert \rightarrow\infty$.

\begin{picture}(300,200)(-60,0)
\put (50,100){\vector(1,0){200}}\put(150,40){\vector(0,1){130}}
\thicklines\put(70,140){\line(1,0){80}}\put(150,140){\line(0,-1){40}}
\put(150,100){\line(1,0){80}}\put (157,136){$i\, T$}\put(155,166){Im~$t$}
\put(65,145){$C_T$}\put(240,87){Re~$t$}\put(135,20){Fig. 1}
\end{picture}

The semiclassical calculation of the transition probability between arbitrary
multiparticle states
above neighboring topological gauge vacua is quite a formidable task.
It requires solving the Yang-Mills equations on a suitably chosen complex time
contour,
with arbitrary boundary conditions imposed at the initial and final times.
In order to simplify the problem,
we will make use of the conformal symmetry of the classical Yang-Mills action
and
reduce the number of degrees of freedom to one.

 Although such a drastic simplification will lead us astray from the problem of
fermion number violation
in high-energy  collisions,
it has the advantage of being tractable analytically and
provides new insight into the role of complex time singularities.
We find that they entirely determine the topological charge and
imaginary part of the action of the solution.
The imaginary part of the action enters the WKB-exponent for the transition
probability between
the initial and final states.
The topological charge, through the anomaly equation,
is the quantity determining the amount of fermion number or
chirality violation in the process.

We consider the SO(4) conformally invariant Minkowski time
($t\in {\bf R}$ ) solutions of L\"uscher and Schechter \cite{LUS,SCH},
analytically continued to a complex time contour (fig. 1).
Our aim is the computation of many-particle transition amplitudes in the
one-instanton sector.
Therefore,
we will consider only a subclass of these solutions which have integer
topological charge on
the complex time contour $C_T$.
Only the solutions with a turning point at say, $t=0$ for all $\vec{x}$,
have this property, as will be made clear at the end of this section.
The \ls solutions are real in Minkowski time.
The turning point condition assures that their analytic continuation to
the Euclidean time axis is real as well.
Note that in general the fields will be complex on the ${\rm Re}\, t<0$
part of the contour,
since $t=iT$ is {\it not} a turning point of the
solution\footnote{It is easy to show that the SO(4)-conformally invariant
solutions can
have at most one turning point.}.

In this section,
we will work in Euclidean time and find a real solution with a turning point at
zero Euclidean time.
Its analytic continuation to Minkowski time will then be real as well.
We define the Euclidean action of  SU(2) pure Yang-Mills theory
to be imaginary for a real Euclidean solution:
\beq
\label{a1}
S \: = \: {i\over 4 g^2}\,\int d^4 x \,  F_{\mu\nu}^a F_{\mu\nu}^a .
\eeq

In order to  make use of the conformal symmetry of the action (\ref{a1}),
it is convenient to introduce new variables,
which simplify the action of the conformal group.
(For details, see \cite{LUS, SCH, FKS}.)
The spatial radius $r \equiv \vert\vec{x}\vert $ and Euclidean time
$\tau = i\, t$ are mapped into two parameters of the Lobachevski plane
$\left(\, w, \phi\,\right)$:
$$
\left\{\, 0\le r < \infty , \: -\infty  < \tau < \infty \,\right\}
\,\longrightarrow\,
\left\{\, -\pi  /2 \le w \le +\pi /2 , \: -\infty  < \phi < \infty \,\right\}
$$
according to the relations:
\beq
\label{map}
\tan w = {r^2 + \tau^2 -1 \over 2r } , \hspace{4mm}
\cosh \phi = { 1 + r^2 + \tau^2 \over 2r } \cos w , \hspace{4mm}
\sinh \phi = {\tau \over r } \cos w \, .
\eeq
The following Jacobian relation holds:
$$
{dr \, d\tau \over r^2} \: = \: {dw \, d\phi \over {\rm cos}^2 w } \, .
$$

L\"uscher and Schechter have shown that the most general
solution for which a $SO(4)$-conformal transformation
can be compensated by a global $SU(2)$-gauge transformation
is parameterized by a single function  $q(\phi)$.
Its action (\ref{a1}) is \cite{LUS, SCH}:
\beqa
\label{a2}
S & = &
i\, {12\pi \over g^2}\int^{\infty}_{-\infty}d \phi
\int_{-\pi /2}^{+\pi /2}dw \:
\cos^2 w \,
\left[\,
{1 \over 2}\,\dot{q}^2 \, +\,{1 \over 2}\, \left(q^2 - 1\right)^2
\,\right] \nonumber \\
& = &
i\, {12\pi \over g^2}\int^{\infty}_0 d r
\int_{-\infty}^{+\infty}d\tau \:
{\cos^4 w \over r^2} \,
\left[\,
{1 \over 2}\,\dot{q}^2 \, +\,{1 \over 2}\, \left(q^2 - 1\right)^2
\,\right] \, ,
\eeqa
where $\dot{q} \equiv {d \over d\phi} q(\phi)$.

The topological charge in terms of the \ls Ansatz becomes :
\beqa
\label{q1}
Q & \equiv &
{1 \over 32\pi^2}\int d^4 x \: F_{\mu\nu}^a \tilde{F}_{\mu\nu}^a
\: = \: {1 \over 2\pi} \int^{\infty}_{-\infty} d\phi
\int_{-\pi /2}^{+\pi /2}dw \:
\cos^2 w \, {d \over d\phi}(q^3 - 3q) \nonumber \\
& = &
{1 \over 2\pi} \int\limits^{\infty}_0 dr
\int\limits_{-\infty}^{+\infty}d\tau \:
{\cos^4 w \over r^2}\,\dot{q}\, \left(\, 3q^2 \, -\, 3 \,\right) \, .
\eeqa

It follows from (\ref{a2}) that the equation of motion for
the Euclidean \ls Ansatz is:
\beq
\label{eom}
\ddot{q} \: = \: - {d \over dq}
\left[\, -{1\over 2}\, \left(\, q^2 - 1 \,\right)^2\,\right] \, ,
\eeq
so that $q(\phi)$ is the coordinate as a function of ``time''$\phi$
of a particle moving in an inverted double-well potential
\beq
\label{pot}
V(q) \: = \: -{1\over 2}\, \left(\, q^2 - 1 \,\right)^2 \, .
\eeq
 Two extrema of the double-well, $q=1$ and $q=-1$,
correspond to vanishing field strengths, $F_{\mu\nu}$\footnote{See (\ref{a-q})
for explicit formulae relating $q$ to the gauge potentials.}.
The respective gauge potentials however differ by a large gauge
transformation with winding number one.

Let us note that this Ansatz contains the BPST instanton \cite{BPST}.
It is given by the solution of (\ref{eom})
$$
q(\phi) \: = \: -\tanh\phi \, ,
$$
representing the motion of a particle which begins at $q = 1$ at time
$\phi=-\infty$ and reaches $q=-1$ at time $\phi=+\infty$.
It is easy to verify using (\ref{a2}) that the action is $8\pi^2 /g^2$
and the topological charge (\ref{q1}) is unity.

As stated previously, we will look only for solutions with
a turning point at say,
$\tau =0$ for all $\vec{x}$.
It follows from the explicit form of the mapping (\ref{map}) that $\tau =0$
is equivalent to $\phi = 0$ for all $r$.
One can then easily verify,
using the explicit formulae relating $q$ to gauge potentials (\ref{a-q}),
that the condition $\dot{q}(\phi = 0)=0$ corresponds to a turning point
of the gauge potentials at $\tau = 0$.
A turning point of the gauge potentials requires $A_0^a (\tau =0, \vec{x})=0$
and $\partial_{\tau} A_i^a (\tau =0, \vec{x})=0 $.
Hence,
the continuation of the gauge potentials to Minkowski time will be real
as well.

A solution of (\ref{eom}) with such a turning point is easy to find
by considering the one-dimensional double-well problem.
It represents oscillatory motion in the well between $q=1$ and $q=-1$
of the potential $V(q)$ (\ref{pot}).
The turning point condition at $\phi = 0$ leaves one free parameter:
the ``energy'' $\epsilon$ ($\epsilon < 1/2$),
or equivalently the initial coordinate, $q_- = \sqrt{1-\sqrt{2\epsilon}}$,
of the particle in the well.
This solution is explicitly given in terms of the Jacobian elliptic
sine\footnote{ The solution (\ref{soltn}) can be shown to equal one of
the Minkowski solutions given in \cite{LUS,FKS} by analytic continuation
to Minkowski time, shift by half a period and use of the transformation
formulae for elliptic functions \cite{GR}.}:
\beq
\label{soltn}
q\left(\,\phi (r, \tau)\,\right) \: = \:
q_{-}\, {\rm sn} \left(\, q_+\phi (r, \tau) \, + \, K \, , \, k\,\right) \, .
\eeq
The two turning points in the well are:
$$
q_{\pm} \: = \: \sqrt{1\pm\sqrt{2\epsilon}} \, ,
$$
and the modulus and primed-modulus of the elliptic sine are:
$$
k^2 \: = \: {q_-^2 \over q_+^2}
\: = \: {1-\sqrt{2\epsilon} \over 1+\sqrt{2\epsilon}} \, ,\hspace{1cm}
k^{\prime 2} \: \equiv \: 1 - k^2 \: = \:
{ 2\sqrt{2\epsilon} \over 1+\sqrt{2\epsilon}} \, .
$$

We shall be interested in what follows in the limit of
small $\epsilon$ (or $k'\rightarrow 0$).
This limit corresponds to solutions which are close to the vacuum
$\vert q\vert =1$ at the turning point.
These solutions are of interest because they will be shown to
describe transitions between initial and final coherent states
containing a different number of particles \cite{RST}.
In this limit,
the periods of the elliptic sine have the following expansion \cite{GR}:
\beq
\label{k}
K \: = \: \ln {4\over k'} \, + \, {k'^2  \over 4}
\left( \, \ln{4\over k'} - 1\,\right) \, +
\, O\left(\, k'^4 \ln k'\,\right)\, , \hspace{1cm}
K' \: = \: {\pi \over 2}\, \left(\, 1 + {k^{\prime 2} \over 4}\,\right) \, +
\, O\left(\,k'^4\,\right) \, .
\eeq
Let us now turn to the calculation of the action and topological
charge of the solution (\ref{soltn}) on the complex time contour $C_T$.

Both the imaginary part of the action and the topological charge are determined
by the singularities of the solutions in the complex time plane,
as will be clearly demonstrated below.
The elliptic sine has only simple poles, whenever \cite{GR}
$$
q_+ \phi \, + \, K \: = \: 2nK \, + \, (2m+1)\, i\, K' \, , \hspace{1cm}
n,m \: \in \: {\bf Z} \, .
$$
Note that when analytically continued to complex $\tau$,
the imaginary part of $\phi$ (\ref{map}) obeys
$\vert{\rm Im}\,\phi\vert\le\pi$.
Therefore,
solutions of the above equation exist in the limit of small $\epsilon$
only for $m=-1$ and $m=0$.
In $\left(r, \tau\right)$-space,
the singularities lie on the curves :
\beq
\label{sing}
\tau_{nm}(r) \: = \: q_{nm} \, \pm \, \sqrt{q_{nm}^2 - 1 - r^2} \, ,
\eeq
where
$$
q_{nm} \: = \: \coth\, {(2n-1)K \, + \, (2m+1)\,i\, K^{\prime} \over q_+} \, .
$$
There is also a ``cross'' of essential
singularities of the mapping (\ref{map})
$\left(r,\tau\right)\rightarrow\left(w,\phi\right)$:
$$
\tau \: = \: \pm (1 \pm ir) \, , \hspace{.3cm} r \neq 0 \, ,
$$
where all four combinations of signs are allowed.
The complex time contour $C_T$ in fig. 1 should therefore be required to
have ${\rm Re}\,\tau < 1$ in order to  avoid them.
The equations of the first two singularity lines (\ref{sing}),
for small $\epsilon$ and $r >> 1$, are:
$$
\tau_{1,-1} (r) \: = \: 1 - 2{\sqrt{2\epsilon} \over 8} - ir \, ,\hspace{1cm}
\tau_{2,-1} (r) \: = \: 1 - 2({\sqrt{2\epsilon} \over 8})^3 - ir \, ,
$$
after making use of the expansions (\ref{k}).
Note that $m=-1$ and $m=0$ correspond to complex conjugation of $\phi$.
Since $\phi$ is real for ${\rm Im}\,\tau = 0$,
it takes complex conjugate values at points with ${\rm Im}\,\tau \neq 0$,
which are reflections of each other with respect to the Euclidean time axis.
Hence,
only the singularity lines with $m=-1$ lie in the ${\rm Im}\,\tau = t <0$
half-plane and are relevant to calculations on the contour $C_T$ in fig. 1.

These singularities are illustrated in fig. 2.
As mentioned above, the parameter $T$ should be chosen to obey
\beq
\label{nu}
{\rm Re}\,\tau_{1,-1} (\infty) \, < \,  T \, < \,
{\rm Re}\,\tau_{2,-1} (\infty)
\eeq
in order to avoid the singularity lines.
Consider now the action (\ref{a2}) on the closed contour $C_T + C_M$,
where $C_M$ runs along the Minkowski time axis for $-\infty < t <\infty$,
and $C_T$ is the contour described above:
\vspace{1mm}
\beq
\label{sct}
S_{C_T} \, + \, S_{C_M} \: = \:
i\, {12\pi \over g^2}\int\limits^{\infty}_0 dr \:
2\pi i \, \sum_{nm} {\rm Res}\,
\left\{\, {\cos^4 w \over r^2}\,
\left[\,
{1 \over 2}\,\dot{q}^2 \, +\,{1 \over 2}\, \left(q^2 - 1\right)^2
\,\right]
\,\right\}
\rule[-3mm]{0.2mm}{9mm}_{\,\tau_{nm}(r)} \, .
\eeq
\vspace{1mm}
The sum is over the singularity lines between the contour $C_T$ and the
Minkowski time axis; the sum contains only one term for the contour in fig. 2.

\begin{picture}(300,200)(-60,0)
\put (50,100){\vector(1,0){200}}\put(150,40){\vector(0,1){130}}
\thicklines\put(70,140){\line(1,0){80}}\put(150,140){\line(0,-1){40}}
\put(150,100){\line(1,0){80}}\put(100,125){\circle*{5}}
\put(110,125){$\tau_{1,-1}$}\put(100,148){\circle*{5}}
\put(110,148){$\tau_{2,-1}$}\put(65,145){$C_T$}\put(157,135){$T$}
\put(155,165){Re~$\tau =$ Im~$t$}\put(240,87){Im~$\tau =$ Re~$t$}
\put (135,20){Fig. 2}
\end{picture}

Now,
the imaginary part of $S_{C_T}$ is the quantity entering the WKB-exponent
of a transition probability dominated by this solution.
Since the solution is real on $C_M$,
the contribution to the action from $C_M$ is purely real.
So, the residue alone determines ${\rm Im}\, S_{C_T}$ :
\vspace{1mm}
\beq
\label{ims}
{\rm Im}\, S_{C_T} \: = \:
 -{24\pi^2 \over g^2}\int^{\infty}_0 dr \:
{\rm Im} \sum_{nm} {\rm Res}\,
\left\{\, {\cos^4 w \over r^2}\,
\left[\,
{1 \over 2}\,\dot{q}^2 \, +\,{1 \over 2}\, \left(q^2 - 1\right)^2
\,\right]
\,\right\}
\rule[-3mm]{0.2mm}{9mm}_{\,\tau_{nm}(r)} \, .
\eeq

Consider now the topological charge $Q$ (\ref{q1}) on the contour $C_T + C_M$.
Our solution (\ref{soltn}) is
an even function of $\phi$, therefore the integral for $Q$ on the Minkowski
time axis vanishes.
Thus,
$Q$ on $C_T$ is determined by the residues at the singularity lines
(\ref{sing}) as well:
\vspace{1mm}
$$
Q \: = \:  3\, i\,\int^{\infty}_0 dr\:
\sum_{nm} \, {\rm Res}\,
\left\{\,
{\cos^4 w \over r^2}\, \dot{q} \, \left(q^2 - 1\right)
\,\right\}\rule[-3mm]{0.2mm}{9mm}_{\,\tau_{nm}(r)} \, .
$$
\vspace{1mm}
Let us concentrate for simplicity on the case when our contour encloses only
one singularity line, as illustrated in fig. 2.
Using the Laurent expansion of the elliptic sine \cite{GR}
at the pole at $\phi_{1,-1} = \coth^{-1} q_{1,-1}$, we find for
the action
\beqa
\label{ims2}
\lefteqn{{\rm Im}\, S_{C_T} \: = \:
-{24\pi^2 \over g^2}\int^\infty_0 dr \:\times} \\
& &
{\rm Im}\, {\rm Res} \, \left\{ {\cos^4 w(r, \tau) \over r^2} \,
\left[\, {1 \over \left(\,\phi (\tau,r) - \phi_{1,-1}\,\right)^4} \:- \:
{2 \over 3}\, {1\over \left(\,\phi (\tau,r) - \phi_{1,-1}\,\right)^2}
\: + \: \cdots  \,\right] \,\right\}
\rule[-3mm]{0.2mm}{9mm}_{\,\tau_{1,-1}(r)} \, , \nonumber
\eeqa
and for the topological charge
\beqa
\label{qresult}
\lefteqn{Q \: = \: 3i\, \int^{\infty}_0 dr\:\times} \\
& & {\rm Res}\,\left\{\, {\cos^4 w(r, \tau) \over r^2}\,
\left[\, {1 \over \left(\,\phi (\tau,r) - \phi_{1,-1}\,\right)^4}
\: - \:
{2 \over 3}\, {1\over \left(\,\phi (\tau,r) - \phi_{1,-1}\,\right)^2}
\: + \: \cdots \right]\,\right\}
\rule[-3mm]{0.2mm}{9mm}_{\,\tau_{1,-1}(r)} \, , \nonumber
\eeqa
where the ellipsis denotes terms regular  as $\phi\rightarrow \phi_{1,-1}$.
Note that the singular terms of the Laurent expansion for ${\rm Im}\, S$
and $Q$ are equal; the regular terms differ, however.
Calculating the residue, we find:
$$
\int^{\infty}_0 dr \:
{\rm Res} \,
\left\{\, {\cos^4 w(r, \tau) \over r^2}
\left[\, {1 \over \left(\,\phi (\tau,r) - \phi_{1,-1} \,\right)^4}
\, - \,
{2 \over 3}\, {1\over \left(\, \phi (\tau,r) - \phi_{1,-1}\,\right)^2} + ...
\right]\, \right\}
\rule[-3mm]{0.2mm}{9mm}_{\tau_{1,-1}(r)} \, , \nonumber
$$
\beqa
\label{res}
& = & -i \, {5 \over 2} \, \int_0^{\infty} dr \:
{ \left(\, q_{1,-1}^2 - 1 \,\right)^2 \, r^2 \,
\sqrt{r^2 + 1 - q_{1,-1}^2} \over
\left(\, 1 \, - \,  q_{1,-1}^2 \, + \, r^2\,\right)^4 } \\
& = & -i \, {5\over 2} \, \int_0^{\infty}
{y^2 \, dy \over \left(\, y^2 \, + \, 1 \,\right)^{7/2}} \: = \: -{i\over 3}
\nonumber \, .
\eeqa
Therefore, we have found that the imaginary part of the action and the
topological charge are:
\beq
\label{action}
{\rm Im}\, S \: = \: {8\pi^2 \over g^2} \, , \hspace{.5cm} Q \: = \: 1.
\eeq
Although the residue for a given $r$ depends on the number $n,m$
of the singularity line, the integral over $r$ does not.
We have shown that these solutions have
 the remarkable property that the
imaginary part of the action and the topological charge obey
\beq
\label{bogom}
{g^2 \over 8\pi^2}\, {\rm Im}\, S \: = \:  Q \: = \: N  \, ,
\eeq
where $N$ is the number of singularity lines between the
complex time contour and the Minkowski time axis.
Of course, this relation is identical to that obeyed by Euclidean
multi-instanton configurations \cite{JNR,WIT}.

It should be stressed that the relation (\ref{bogom}) is
far from trivial on the complex time contour.
The usual arguments for establishing the
Bogomol'nyi bound  do not seem to hold here,
since the fields take complex values on the contour \cite{BOG}.
The turning point condition at $\tau =0$ is crucial for (\ref{bogom}) to hold.
As was shown in \cite{FKS},
the Minkowski time topological charge vanishes only for solutions
with a turning point.
The \ls\hspace{.2cm}solutions without a turning point have fractional
topological
charge on the contour $C_T$\footnote{The charge on the contour $C_T$ is
in this case the sum of an (integer) residue and a (fractional \cite{FKS})
Minkowski time contour contribution.}.

Let us also note that the real part of the action
of our solution on $C_T$  coincides, up to a minus sign, with the
action on the Minkowski time contour $C_M$.  This follows from the fact
that the residue (\ref{res}) is purely imaginary.  As was shown in
\cite{LUS}, the action and the energy (See (\ref{energy}).) of the purely
Minkowski solution are also finite.

\mysection{Free Wave Asymptotics of the Gauge Field}

In this section,
we will find the free-wave asymptotics of the solution at the initial
and final times, which determine the initial and final coherent states.
We will explicitly demonstrate that the gauge field asymptotics at
$t\rightarrow +\infty$ and $t \rightarrow -\infty$ belong to different
topological sectors, confirming  the calculation of the topological charge
of the previous section.

The complex time contour $C_T$ provides a natural way of incorporating
non-vacuum boundary conditions at the initial and final times.
In the semiclassical approximation, the initial and final states are coherent
states of the form:
\beq
\label{cohstate}
\vert\, \{ d ({\bf k}) \} \,\rangle \: = \:
\exp\left[ \, \int d {\bf k}\, d({\bf k})\, \hat{a}^{\dagger} ({\bf k})\,
\right]   \,\vert\, 0\,\rangle
\eeq
The creation operator is $\hat{a}^{\dagger}({\bf k})$ and all color and
polarization indices have been suppressed.
The complex amplitudes $d({\bf k})$ are determined by the free-field
asymptotics of the solution at the ends of the contour $C_T$, in a manner to be
discussed in the next section.
In order to find them we need to know the Fourier transforms of the gauge
potentials at the initial and final times.

Our strategy will be to start from the Minkowski part of the contour,
at $t\rightarrow\infty$, and find the field asymptotics determining the final
coherent state. Considering then the analytic properties of the solution
(\ref{soltn}) in the complex-$r$ plane, we will establish a simple relation
between the Fourier transforms at the initial and final times, and will thus be
able to determine the initial coherent state as well.

To begin, we note the formulae relating the solution of the
one dimensional problem (\ref{eom})
and the four dimensional gauge potentials \cite{LUS,SCH}.
In Minkowski space with metric $g_{\mu\nu}=(-,+,+,+)$,
the gauge potentials are expressed through
the solution $q (r, it)$ (\ref{soltn}) as:
\beq
A_0^a \,\left(\, {\bf r}, t\,\right) \: = \:
{4\over g}\, \left( q(r,it) - 1 \right)\,
{t \, r_a \over \left( r^2 + 1 - t^2 \right)^2 + 4t^2} \, ,
\eeq
\beq
\label{a-q}
A_l^a \,\left(\, {\bf r}, t\,\right) \: = \:
- {4\over g}\, \left( q(r,it) - 1 \right)\:
{ {1\over 2}(1+t^2 -r^2)\delta_{al}
\, + \, \epsilon_{alm}r_m \, + \, r_a r_l \over
\left(r^2 + 1 - t^2\right)^2 \, + \, 4t^2 }  \, .
\eeq
Let us define a function
\beq
\label{p}
P(k,t) \: \equiv \: {4 \over g}\, \int d^3 r \:
{e^{i{\bf kr}}\, \left(q(r,it) - 1\right) \over
\left(r^2 + 1 - t^2\right)^2 + 4t^2}
\: = \: {8\pi \over ikg}\,\int_{-\infty}^{\infty}
dr \: {r\, e^{ikr} \, \left(q(r,it) - 1\right) \over
\left(r^2 + 1 - t^2\right)^2 \, + \, 4t^2 } \, .
\eeq
Then, the Fourier transforms of the gauge potentials (\ref{a-q}) are easily
expressed in terms of $P(k,t)$ as:
\beq
\label{a-p}
A_0^a\, \left( {\bf k}, t\right) \: = \:
-i{k^a\over k}\, t \, {\partial P\over \partial k} \, ,
\eeq
\beq
\label{a-p2}
A_i^a \, \left({\bf k}, t\right) \: = \:
-\delta_{ai}\left(
{1\over 2}(1 + t^2) P \, + \, {1\over 2}{\partial^2 P\over \partial k^2}
\right) \: + \:
i\epsilon_{aim}{k_m \over k} {\partial P\over \partial k} \: + \:
{k_i k_a \over k^2}\left({\partial^2 P\over \partial k^2} -{1\over k}{\partial
P\over \partial k}\right) \, .
\eeq

Recall that a  pure gauge configuration with unit winding number, as explained
in the previous section,
is given by the potentials (\ref{a-q}) with $q=-1$, the
second extremum of the double-well potential $V(q)$ (\ref{pot}).
For further use,
let us denote the function (\ref{p}), corresponding to this configuration
by $\pi(k,t)$:
\beq
\label{pi}
\pi \left(k,t\right) \: = \: {4\pi^2 i \over gkt}\, e^{ikt -k} \: - \:
{4\pi^2 i \over gkt}\, e^{-ikt -k}.
\eeq

In order to calculate the Fourier transforms of the gauge fields at
large Minkowski time,
we note that at large $t$ the solution (\ref{soltn}) represents
a thin shell of energy, expanding with the speed of light (see Fig. 3.).

For this configuration,
the surface energy density decreases like $1/r^2\sim 1/t^2$ and we expect
the nonlinear terms to become subdominant in the infinite time limit.
Hence, as $t\rightarrow\infty$,
the solution reduces to a solution of the free equations of motion.
{}From the Fourier series expansion of the elliptic sine \cite{GR},
we see that the only terms which solve the free equations are
those proportional to $1$ and $\cos (2 i \phi (r, it))$.
(Note that free equations for $A_{\mu}^a$ correspond to a harmonic
approximation for $q$ around one of the minima of the double well
(\ref{pot}).).
Therefore, in the limit of large time,
the solution (\ref{soltn}) has the following representation:
\beq
\label{solution}
q(r,it) \: = \:  1 \, - \, \sqrt{{\epsilon \over 2}}\, \cos
2 \, i \phi (r, it)
\: = \:
1 \, - \, \sqrt{{\epsilon \over 2}} \,
{ \left(r^2 + 1 - t^2\right)^2 \, - \,  4t^2 \over
\left(r^2 + 1 - t^2\right)^2 \, + \,  4t^2}  \, ,
\eeq
The coefficient in front of the second term is fixed by the
requirement that the energy at infinite time equals the exact energy
of the classical solution.  (See (\ref{energy}) and fig. 3.)
The corresponding function $P_{\rm fin}(k,t)$ is:
\beq
\label{init}
P_{{\rm fin}}(k,t) \: = \:
{\pi^2 \, i \, \sqrt{2\epsilon} \over g\, (t-i)}\, e^{-ikt - k} \: - \:
{\pi^2 \, i \, \sqrt{2\epsilon} \over g\, (t+i)}\, e^{ikt -k } \, .
\eeq

The Fourier transforms of the gauge potentials are then obtained in the form:
\beq
\label{fp}
A_0^a \, ({\bf k},t) \: = \:
- i\, {\pi^2\sqrt{2\epsilon} \over g}\, {k^a \over k}\, t \, e^{ikt} \, e^{-k}
\: + \:
\left\{ {\rm h.c.} \hspace{.1cm} {\rm and} \hspace{.1cm}{\bf k}\rightarrow
-{\bf k}\right\}  \, ,
\eeq
\beqa
\label{fp2}
A_l^a\, ({\bf k}, t) & = & {\pi^2\sqrt{2\epsilon}\over g}\,
\left[\,
\delta_{al} \, e^{ikt} \, e^{-k} \: + \:
i\, {\epsilon_{alm}k_m \over k}\, e^{ikt} \, e^{-k} \right. \\
&   & \hspace{2.5cm} \left.
+ \: {k_a k_l \over k^2}\, \left(it  - 1 - {1 \over  k} \right)
\, e^{ikt} \, e^{-k} \: + \:
\left\{\,{\rm h.c.} \hspace{.1cm} {\rm and} \hspace{.1cm}{\bf k}\rightarrow
-{\bfk}\,\right\}
\,\right] \, . \nonumber
\eeqa
In order to see that they indeed obey the free equations of motion,
it is convenient to represent them as a purely transverse part
plus Abelian pure gauge\footnote{For a discussion of the asymptotic behaviour
of classical Yang-Mills solutions in Minkowski space, see L\"uscher
\cite{LUS2}. The ``radiation data'' (\ref{afin}) are sufficient to determine
the momentum distribution of the outgoing waves.}:
\beq
\label{afin}
A_0^{a }({\bf k},t)\: = \: {\partial \over \partial t} \,
\omega^a ({\bf k},t) \, ,
\eeq
\beqa
\label{afin2}
A_l^{a }\, ({\bf k},t) & = & {\pi^2\sqrt{2\epsilon}\over g}\,
\left[\, \left(\delta_{al} -{k_a k_l \over k^2}\right) \, e^{ikt} \, e^{-k}
\: + \: i \, {\epsilon_{alm}k_m \over k}\, e^{ikt}\, e^{-k} \right. \\
&   & \left. \hspace{3.5cm} \rule{0mm}{6mm}
+ \: \left\{\,
{\rm h.c.} \hspace{.1cm} {\rm and} \hspace{.1cm}{\bf k}\rightarrow -{\bf k}
\,\right\}
\right] \: - \: i\, k_l\, \omega^a ({\bf k},t) \nonumber \\
& = & {1\over \sqrt{2k}} \, \sum_{i=1}^2 \left[\,
e^i_l ({\bf k}) \, g^{a *}_i ({\bf k})\, e^{ikt} \: + \:
e^i_l (-{\bf k})\, g^{a}_i (-{\bf k}) \, e^{-ikt} \,\right]
\: - \: i\, k_l\, \omega^a ({\bf k},t)  \, ,
\eeqa
with the Abelian gauge function
\beq
\label{fingauge}
\omega^a ({\bf k},t) \: =  \:
i\, {\pi^2\sqrt{2\epsilon}\over g}\, {k^a \over k^2} \,
\left(\, it \, - \, {1\over k} \,\right)
\, e^{ikt} \, e^{-k} \: + \:
\left\{\,
{\rm h.c.} \hspace{.1cm} {\rm and} \hspace{.1cm}{\bf k}\rightarrow -{\bf k}
\,\right\} \,  .
\eeq
Here $ e^i_l ({\bf k})$ are the two transverse polarization vectors,
obeying $e^i_l \, e^i_m = \delta_{lm} - k_l k_m/k^2$,
and
\beq
\label{b}
g^{a }_i \, ({\bf k}) \: = \: {\pi^2\sqrt{2\epsilon}\over g}\,\sqrt{2k}\,
e^{-k}\, \left(\,
e^i_a ({\bf k}) \, - \, i\, {\epsilon_{alm}k_m \over k} \, e^i_l ({\bf k})
\,\right) \, ,
\eeq
\beq
\label{b2}
g^{a *}_i ({\bf k}) \: = \: \left[\, g^{a}_i ({\bf k})\,\right]^*.
\eeq

The calculation of the Fourier transforms of the fields at initial time,
on the complex part of the contour $C_T$, is less straightforward.
Since  at large early times the contour is trapped between two
singularity lines (\ref{sing}),
the approximation we used for the solution in Minkowski time cannot
be justified.
However,
consideration of the analytic properties of the solution (\ref{soltn})
in the complex-$r$ plane will allow us to relate the field asymptotics
at the initial times to those at final times.

For the initial state,
the formulae (\ref{a-q}), (\ref{p}), (\ref{a-p}) are applicable as well,
up to the replacement $t \rightarrow t+iT$.
Hence, we need to calculate the function
\vspace{1mm}
\beq
\label{pin}
P_{{\rm in}}(k,t+iT) \: = \: {8\pi \over ikg}\, \int_{-\infty}^{\infty}
dr\: {r\, e^{ikr} \, \left(\, q(r,i(t+iT)) - 1 \,\right) \over
\left(r^2 + 1 - (t+iT)^2\right)^2 \, + \, 4(t+iT)^2 } \, .
\eeq
\vspace{1mm}
The corresponding function  for the final state is
\vspace{1mm}
\beq
\label{pfin}
P_{{\rm fin}}(k,t) \: = \: {8\pi \over ikg}\, \int_{-\infty}^{\infty}
dr\: {r\, e^{ikr} \, \left(\, q(r,it) - 1 \,\right) \over
\left(r^2 + 1 - t^2\right)^2 \, + \, 4t^2} \, .
\eeq

If no poles of the integrand in (\ref{pfin}) crossed the real-$r$ axis
when analytically continued to $t+iT$, we would have
$P_{{\rm in}}(k,t+iT) \, = \, P_{{\rm fin}}(k,t+iT)$. The Fourier
transform of the gauge field for the initial state would  then be given
by (\ref{afin}), with the replacement $t\rightarrow t+iT$, and the  initial and
final  coherent states would be the same.

However, this is not the case for our contour $C_T$. The solution
$q(r, it)$ (\ref{soltn}) has poles in the complex-$r$ plane, the
positions of which are given by the inversion of the equation of the
singularity lines (\ref{sing}). Let us define $T\equiv 1-\nu$, where,
according to (\ref{nu})
\vspace{1mm}
\beq
\label{nu1}
2({\sqrt{2\epsilon} \over 8})^3 \: < \: \nu \: < \:
2{\sqrt{2\epsilon} \over 8} \, .
\eeq
\vspace{1mm}
In this case, it is easy to see that when continued from $t$ to
$t+iT$, exactly two poles of the integrand in (\ref{pfin}) cross the
real-$r$ axis. The pole at
\beq
\label{pole1}
r_0^- (t) \: = \: -t \, + \, i \, -\, 2\, i\,\alpha  \, , \hspace{2cm}
\alpha \:\equiv\:{\sqrt{2\epsilon}\over 8} \, - \, {\epsilon\over
16}\,\ln\epsilon \, ,
\eeq
crosses the real axis from above, and the one at
\beq
\label{pole2}
r_0^+ (t) = t - i +  2\, i\,\alpha
\eeq
crosses  from below,
pushing thus the integration contour in (\ref{pfin}) off the real axis.
Hence, $P_{{\rm fin}}(k, t+iT)$ is given by
\beq
\label{pfinal}
P_{{\rm fin}}(k,t+iT) \: = \: {8\pi \over ikg}\int_{C_I}
dr \: {r\, e^{ikr}\, \left(\, q(r,i(t+iT)) - 1 \,\right) \over
\left(r^2 + 1 - (t+iT)^2\right)^2 \, + \, 4(t+iT)^2} \, ,
\eeq
where the contour $C_I$ is shown in fig. 4.

\begin{picture}(300,200)(-60,0)
\put (50,100){\vector(1,0){200}}\put (90,100){\vector(-1,0){0}}
\put (210,100){\vector(-1,0){0}}\put (150,40){\vector(0,1){130}}
\put (90,90) {\circle*{5}}\put (90,90) {\oval(20,20)[b]}
\put (50,90){\line(1,0){30}}\put (70,90){\vector(1,0){0}}
\put (100,90){\line(5,1){100}}\put (125,95){\vector(4,1){0}}
\put (175,105){\vector(4,1){0}}\put (210,110){\circle*{5}}
\put (210,110){\oval(20,20)[t]}\put (220,110){\line(1,0){30}}
\put (240,110){\vector(1,0){0}}\put (50,78){$C_I$}
\put (155,165){Im~$r$} \put (240,87){Re~$r$}\put (135,20){Fig. 4}
 \end{picture}

Cauchy's theorem then gives:
\setlength{\jot}{3mm}\beqa
\label{in-f}
\lefteqn{P_{{\rm in}}(k,t+iT) \: = \: P_{{\rm fin}}(k,t+iT) \; - \:} \\
& &  2\,\pi i \,\left[\,
{\rm Res}\,\rule[-3mm]{0.2mm}{9mm}_{\, r_0^- (t+iT)} \, - \,
{\rm Res}\,\rule[-3mm]{0.2mm}{9mm}_{\, r_0^+ (t+iT)}\,\right]
 \left\{\,
{8\pi \over ikg}\, {r\, e^{ikr}\, q(r,i(t+iT)) \over
\left(r^2 + 1 - (t+iT)^2\right)^2 \, + \, 4(t+iT)^2 } \,\right\} \, . \nonumber
\eeqa
Calculating the residues is straightforward and yields:
\beq
P_{{\rm in}}(k,t+iT) \: = \: P_{{\rm fin}}(k,t+iT) \: - \:
{4\pi^2 i\over gk(t+iT)}\,
\left[\, e^{ikr_0^- (t+iT)} \: - \: e^{ikr_0^+ (t+iT)}\,\right] \, .
\eeq
Substituting the equations for the poles (\ref{pole1}), (\ref{pole2})
we obtain:
\beqa
\label{get}
\lefteqn{P_{{\rm in}}(k,t+iT) \: = \: P_{{\rm fin}}(k,t+iT)
\: + \: \pi(k, t+iT) } \\
& & -\: {8\pi^2 \, i\, e^{-ik(t+iT)} \over gk(t+iT)}\, e^{-k(1 -\alpha)} \,
\sinh k\alpha \: + \:
{8\pi^2\, i\, e^{ik(t+iT)} \over gk(t+iT)}\, e^{-k\alpha} \,
\sinh k(1-\alpha) \, . \nonumber
\eeqa
\vspace{1mm}
In this formula, the function $\pi(k, t+iT)$ (\ref{pi}) corresponds
to a topologically nontrivial vacuum configuration with unit winding number.
Its appearance provides an explicit confirmation of the fact that
the initial and final states belong to different topological sectors.

We expect that this result is quite general.
It is not difficult to show that for a complex time contour enclosing,
say two singularity lines,
a calculation similar to the previous one can be obtained.
In this case,
two poles of the integrand in (\ref{pfin}) will cross the real-$r$
axis from above and two from below.
The $\epsilon$-independent part of the residue at each pole gives a
contribution to $P_{\rm in}$ which corresponds to a topologically nontrivial
vacuum configuration
\footnote{More precisely, the residues appear with alternating signs,
as a result of the fact that the conformally invariant Ansatz only
distinguishes pure gauge configurations with unit difference of topological
charge.}.

Now, after removing  the $\pi(k, t+iT)$ piece by a large gauge transformation,
and substituting our expression (\ref{init}) for $P_{{\rm fin}}(k,t+iT)$,
we obtain for the negative frequency part of $P_{\rm in}$:
\beq
P_{{\rm in}}^- (k,t+iT) \: = \:
{i\,\pi^2 \over g}\, e^{-ik(t+iT) - k}\,\left(\,{\sqrt{2\epsilon}\over t-i+iT}
\: - \: { 8\over t+iT}\,{e^{k\alpha}\over k}\,\sinh k\alpha \,\right) \, .
\eeq
Expanding in $\epsilon$, we find
\beqa
P_{{\rm in}}^- (k,t+iT) & = &
{i\,\pi^2 \over g}\, e^{-ik(t+iT) - k}\,\sqrt{2\epsilon}
\left(\, { 1\over t-i+iT} \, - \, { 1\over t+iT} \,\right) \\
&   & \hspace{3cm}
+\: { i\,\pi^2 \over 2g(t+iT)}\,\left(\,\epsilon\,\ln\epsilon
\, - \, {\epsilon\, k\over 2}\,\right)\, e^{-ik(t+iT) - k} \, . \nonumber
\eeqa
Using (\ref{a-p}),
the negative frequency part of the gauge potentials can be represented,
analogously to (\ref{afin}), as:
\beq
A_0^{a -}({\bf k},t+iT) \: = \: {\partial \over \partial (t+iT)}
\,\omega^{a -} ({\bf k},t+iT) \, ,
\eeq
\beqa
\lefteqn{A_l^{a - } ({\bf k},t+iT) \: = \: {i\,\pi^2 \over 2g}\,
\left(\,\epsilon\,\ln\epsilon \, - \,{\epsilon\, k\over 2}\,\right)
\,\epsilon_{alm}{k_m \over k} \, e^{-ik(t+iT)}\, e^{-k} } \\
&  & -\: \left(\,\delta_{al} \, - \, {k_a k_l \over k^2}\,\right)\,
{\pi^2\,\epsilon \over 4g}\, (k-1) \, e^{-ik(t+iT)} \, e^{-k} \: - \:
i\, k_l\,\omega^{a -}({\bf k},t+iT) \, , \nonumber
\eeqa
with the Abelian gauge function
\beq
\label{ingaugeneg}
\omega^{a -}({\bf k},t) \: = \: {\pi^2\,\epsilon \,\ln\epsilon \over 2g}\,
{k^a\over k^2}\,(t -{i \over k})\,e^{-k}\, e^{-ikt} \: - \:
{\pi^2\,\epsilon \over 4g}\,{k_a\over k}\,\left(t - i \right) \, e^{-ikt} \,
e^{-k}  \, .
\eeq

The purely transverse negative frequency part of the gauge field at the initial
time is therefore given by:
\beq
\label{ainneg}
A_l^{a -,{\rm tr}} ({\bf k},t+iT)
\: = \: {1\over \sqrt{2k}}\,
 \sum_{i=1}^2 \: e^i_l (-{\bf k}) \, f^{a}_i (-{\bf k}) \, e^{-ikt}  \, ,
\eeq
with
\beqa
\label{f}
\lefteqn{f^{a }_i ({\bf k}) \: =} \\
& &  {i\,\pi^2 \over 2g}\,\sqrt{2k}\,
\left(\,{\epsilon\, k\over 2} \, + \,\epsilon\ln{1\over \epsilon}\,\right)
\,\epsilon_{alm}\,{k_m \over k} \,
e^i_l ({\bf k})
\, e^{k(T-1)} \: - \:  {\pi^2\,\epsilon \over 4g} \,\sqrt{2k}\, (k-1) \,
e^a_i({\bf k}) \, e^{k(T-1)} \, . \nonumber
\eeqa
The negative frequency components determine the initial coherent state.
The calculation of the positive frequency part of the gauge potentials
proceeds along the same lines, the result being:
\beq
A_0^{a +}({\bf k},t+iT) \: = \: {\partial \over \partial (t+iT)} \,
\omega^{a +} ({\bf k},t+iT) \, ,
\eeq
\beq
\label{ainpos}
A_l^{a + } ({\bf k},t+iT) \: = \: {1\over \sqrt{2k}}\,
\sum_{i=1}^2 \: e^i_l ({\bf k}) \, \bar{f}^a_i ({\bf k}) \, e^{ikt}
\: - \: i\, k_l \,\omega^{a +}({\bf k},t+iT) \, ,
\eeq
\vspace{1mm}
with the Abelian gauge function
\beqa
\label{ingaugepos}
\lefteqn{\omega^{a +}({\bf k},t) \: = } \\
& & {8\pi^2 \over g}\,{k^a\over k^2}
\,\left[\,
(t+i\alpha){{\rm sh}k(1-\alpha )\over k} -
i\,\left({\sinh k(1-\alpha )\over k}\right)' \: + \:
i\, {\sinh k(1-\alpha)\over k^2}
\,\right] \, e^{ikt-\alpha k} \, , \nonumber
\eeqa
and
\beqa
\label{fbar}
\bar{f}^a_i  ({\bf k}) & = & -{8\pi^2 \over g}\, e^i_a ({\bf k}) \,
\left[\, \alpha\, {\sinh k(1-\alpha )\over k}
\: - \: \left(\, {\sinh k(1-\alpha ) \over k} \,\right)'
\,\right] \,\sqrt{2k} \, e^{-kT- \alpha k}  \\
&   & \hspace{2.5cm}
- \: i\, {8\pi^2 \over g} \,\epsilon_{alm}{k_m \over k} \,
e^i_l ({\bf k}) \,{\sinh k(1-\alpha ) \over k}
\,\sqrt{2 k} \, e^{-kT- \alpha k} \, . \nonumber
\eeqa

This completes the calculation of the free field asymptotics of the solution.
They are given by
(\ref{afin}), (\ref{b}) at $t\rightarrow \infty$ and
(\ref{ainneg}), (\ref{f}), (\ref{ainpos}), (\ref{fbar}) at
${\rm Re}\, t\rightarrow -\infty$. We saw explicitly that they belong to
different topological sectors (\ref{get}, \ref{pi}).  We also saw that
they obey the free equations of motion and therefore determine the initial
and final coherent states, as we will show in the next section.

\mysection{Initial and Final States}

In this section,
the role of the gauge field configuration (\ref{soltn},\ref{a-p},\ref{a-p2})
in multiparticle scattering amplitudes will be explained.
The gauge field configuration  will be demonstrated to be the dominant
contribution to an inclusive transition probability from a fixed initial
state, in the saddle point approximation \cite{RSTnew}.
The initial state, and the most probable final state for transition from
this initial state, will be characterized by the asymptotics found in the
previous section.

The total transition probability from an initial coherent state,
$ \vert\,\{ \ak \}\,\rangle $,
projected onto  fixed center-of-mass energy $E$, is:
\beq
\label{eq:probability}
\sigma_E \left( \{ \ak \} \right) \: = \:
\sum_{f} \,\vert\,
\langle\, f \,\vert\,\hat{S}\, P_Q\, P_E\, \vert\, \{ \ak \} \,\rangle\,
\vert^{\, 2}   \, .
\eeq
$P_E$ is a projection operator onto states of fixed center-of-mass energy, $E$.
The probability is unity unless the initial state is projected
also onto a subspace which does not commute with the Hamiltonian;
a projection operator $P_Q$ onto states of fixed winding number $Q$
is implicit in our choice of a classical field with this property.
Furthermore, the inclusive sum is over all final states built above
a neighboring sector of the periodic vacuum.

This quantity is relevant to the study of multiparticle cross-sections
for the following reason.
When summed over all initial states,
\beq
\label{eq:microcanonical}
\sigma_E \: = \: \sum_a\:\sigma_E \left( \{ \ak \} \right)
\eeq
it gives the ``microcanonical'' transition probability in the one-instanton
sector; the probabilities of transition from all states of energy $E$ are
equally weighted in this sum.
When evaluated in the saddle point approximation,
$\sigma_E $ yields the maximal transition probability among all states with
energy $E$.
It sets therefore an upper bound on the two-particle inclusive cross-section
in the one-instanton sector \cite{KRT3}.

The semiclassical approximation to (\ref{eq:probability}) will be
made clear by expressing it in an exponential form.
The S-matrix in the interaction picture is
\beq
\hat{S} \: = \: \lim_{t_{i,f}\rightarrow\mp\infty}
\: e^{i\hat{H}_0 t_f} \: e^{-i\hat{H}(t_f - t_i)} \: e^{-i\hat{H}_0 t_i} \, .
\eeq
Inserting a complete set of eigenstates of the gluon field operator
$A$ at initial and final times, we obtain:
\beqa
\label{eq:probability2}
\lefteqn{\sigma_E \left( \{ \ak \} \right)  \: = } \\
& & \sum_f\: \rule[-3mm]{0.2mm}{9mm} \,
\int  dA_f  dA_i \:
\langle\, f \,\vert\, e^{i\hat{H}_0 t_f} \, \vert A_f \rangle \,
\langle A_f \vert\, e^{-i\hat{H}(t_f - t_i)} \, P_Q \,\vert A_i\rangle \,
\langle A_i\vert\, e^{-i\hat{H}_0 t_i} \, P_E \,\vert\,\{ \ak \}\,\rangle
\,\rule[-3mm]{0.2mm}{9mm}^{\, 2}  \,  . \nonumber
\eeqa

Each of the matrix elements in (\ref{eq:probability2}) may now be written
in exponential form.  The matrix element of the evolution operator between
states of the field operator is the Feynman Path Integral:
\beq
\label{eq:FPI}
\langle\, A_f\,\vert\, e^{-i\hat{H}(t_f - t_i)}\, P_Q\,\vert\, A_i\,\rangle
\: = \:
\int_{A_i}^{A_f}\: \left[{\cal D} A\right]_Q \: \exp\left[ i S \right] \, ,
\eeq
with boundary conditions $A\rightarrow A_{i,f}$ as $t\rightarrow t_{i,f}$,
and the integral being taken over fields with topological charge \mbox{$Q$.}
The matrix element involving the initial state is the wavefunctional of
the initial state.
The projection onto states of fixed energy $E$ may be expressed in an
exponential form as follows:
\beqa
\label{eq:initialwf}
\langle\, A_i\,\vert\, e^{-i\hat{H}_0 t_i} \, P_E \,\vert\,\{ \ak \}\,\rangle
& = &
\int_{-\infty}^{+\infty}  d\xi \: e^{-i E \xi} \,
\langle\, A_i \,\vert\, e^{-i\hat{H}_0 t_i} \,e^{i\xi \hat{H_0}} \,\vert\,\{
\ak \}\,\rangle \\
& = &
\int_{-\infty}^{+\infty}d\xi \: e^{-i E \xi} \,
\langle\, A_i \,\vert\, e^{-i\hat{H}_0 t_i} \,\vert\, \{ \ak e^{i \xi
k}\}\,\rangle  \nonumber \\
& = &
\int\limits_{-\infty}^{+\infty} d\xi \: e^{-i E \xi} \,
\exp{\left(\, B_i [ \ak e^{i \xi k} ,A_f ] \,\right)} \, . \nonumber
\eeqa
The functional $B_i$ depends on the field asymptotics at early times
\beqa
\label{eq:bi}
B_i [  \ak , A_i ]  & = &
-\: {1\over 2}\,\int d{\bf k}\:\ak\, a({\bf -k}) \, e^{-2ikt_i}
\: -\: {1\over 2}\,\int d{\bf k}\: k\, A_i({\bf k})\, A_i(-{\bf k})  \\
&   &  \hspace{1.5cm}
+\:\int d{\bf k}\:\sqrt{2k}\,\ak\, A_i({\bf k}) \, e^{-ikt_i}\nonumber \, ,
\eeqa
where color and polarization indices have been suppressed.
$A_i ({\bf k}) $ is the 3-dimensional Fourier transform of the
field $A$, evaluated at initial time $t_i$.

The matrix element involving the final state may be put in a similar
form by inserting the decomposition of unity in terms of coherent states
$$
\sum_f \: \vert\, f\,\rangle\,\langle\, f\,\vert \: = \:
\int {\cal D}b^*\, {\cal D}b \: e^{-\int d {\bf k}\, \bks\bk}
\: = \: 1 \, .
$$
Then,
the transition probability $\sigma_E \left(\{\ak\}\right)$ becomes:
\beqa
\label{sa}
\lefteqn{\sigma_E \left( \{ \ak \} \right) \: = \:
\int {\cal D}b^*\, {\cal D}b\, e^{-\int d {\bf k}\, \bks\bk} \times} \\
& & \rule[-4mm]{0.2mm}{9mm}\,
 \int  dA_f (x) dA_i (x) \: \langle \{\bks  \}\,\vert \, e^{i\hat{H}_0 t_f}\,
\vert A_f \rangle\,
\langle A_f\vert\, e^{-iH(t_f - t_i)} \, P_Q\,\vert A_i\rangle\,
\langle A_i\vert\, e^{-i\hat{H}_0 t_i}\, P_E\,\vert\{\ak  \} \rangle
\,\rule[-4mm]{0.2mm}{9mm}^{\, 2}  \, . \nonumber
\eeqa
in terms of the wavefunctional of the final state,
\mbox{$\langle \{\bks \} \vert A_f \rangle $.}
This will allow us to resolve the most probable final coherent state
$\vert\,\{\bk\}\,\rangle $ from the inclusive sum.

The wavefunctional of the final coherent state is
\beq
\label{eq:finalwf}
\langle\,\{ \bks \}\,\vert\, e^{i\hat{H}_0 t_f}\,\vert\, A_f\,\rangle  \: = \:
\exp{\left(\, B_f [\bks , A_f] \,\right)}
\eeq
with a functional defined similarly to \mbox{(\ref{eq:bi}):}
\beqa
\label{eq:bf}
B_f [ \bks, A_f ]  & = &
-\: {1\over 2}\,\int d{\bf k}\:\bks\, b^*({\bf -k}) \, e^{2ikt_f}
\: -\: {1\over 2}\,\int d{\bf k}\: k \, A_f({\bf k})\, A_f(-{\bf k})  \\
&   & \hspace{1.5cm}
+\: \int d{\bf k}\:\sqrt{2k}\,\bks\, A_f(-{\bf k}) \, e^{ikt_f} \, .\nonumber
\eeqa
$A_f ({\bf k}) $ is the 3-dimensional Fourier transform of the
field $A$, evaluated at final time $t_f$.

The transition probability is now expressed as a path integral of
an exponential by combining these factors:
\beq
\label{eq:exponential}
\sigma_E \left( \{ a \} \right)  \: =  \:
\int {\cal D}b^*\, {\cal D}b \, d\xi d\xi' \, {\cal D}A \, {\cal D}A' \:\exp{
{\bf W} } \, ,
\eeq
\beqa
\label{eq:exponent}
{\bf W} & = &
-{1\over 2}\int d{\bf k}\:\bks\,\bk \: - \: i\, E\,\xi \: + \:
B_i[ \ak e^{ik\xi} , A_i]  \\
&   & \hspace{1cm} + \: B_f[ \bks  , A_f] \: +\: iS(A)
\: +\:
\left\{\, {\rm h.c.} \hspace{.1cm} {\rm and} \hspace{.1cm} \xi \rightarrow
\xi^{\prime} , \hspace{.1cm} A \rightarrow A'  \,\right\} \, . \nonumber
\eeqa
This integral is dominated by its saddle point value if every
term in the exponent is of order $1/g^2$ as $g\rightarrow 0$.
The saddle point conditions have been derived in \cite{RSTnew}:
\begin{enumerate}

\item Variation of ${\bf W}$ with respect to $A$ and $A'$ requires that the
fields obey the Yang-Mills equations of motion on the complex time contour
$C_T$.
The time contour is chosen so that the topological charge of the saddle point
field configuration \mbox{is $Q$.}
The \ls solution (\ref{soltn}) with a turning point has these properties
on the complex time contour $C_T$  (fig. 2), as we showed in Section 2.

\item Variation with respect to $b_{\bf k}$ and $b^*_{\bf k}$ requires that
$A=A'$ everywhere in space-time.  So, we need only consider a single solution
to the equations of motion.

\item Variation with respect to the initial and final values of the fields
$A_i, A_f$ and $A'_i, A'_f$ relates the saddle point values of
$b_{\bf k}$, $b^*_{\bf k}$ and $a_{\bf k}$, $a^*_{\bf k}$ to the field
asymptotics at $\vert\, t\,\vert\rightarrow\infty$, found in the
previous section. So, variation with respect to the field values at
$t\,\rightarrow \, +\infty$ yields the boundary condition, relating the field
asymptotics at the final time to the complex amplitudes, determining
the most probable final state:
\beq
\label{bcfin}
i\dot{A}_f ({\bf k})\, -\, k\, A_f ({\bf k})\, +\,\sqrt{2k}\, b^*({\bf k})\,
e^{ikt_f}\, =\, 0\, .
\eeq
Here $A_f ({\bf k})$ is the three-dimensional Fourier transform of the
classical solution at the final time, given by equations (\ref{ainneg},
\ref{ainpos}, \ref{afin}).
(Color and polarization indices are suppressed.)

Similarly, the  condition, matching the complex amplitudes of the initial
 coherent state with the solution, is derived by varying   ${\bf W}$ with
respect
to the  values of the fields at ${\rm Re}\, t\,  \rightarrow \, -\infty$ :
\beq
\label{bcin}
-i\dot{A}_i ({\bf k})\, - \, k\, A_i ({\bf k})\, +\, \sqrt{2k}\, a\, (-{\bf k})
\, e^{-ikt_i + ik\xi}\, =\, 0\, .
\eeq
$A_i ({\bf k})$  is the three-dimensional Fourier transform of the classical
solution at the initial time, given by (\ref{ainneg}, \ref{ainpos},  \ref{f},
\ref{fbar}).

\item Variation with respect to $\xi$ and $\xi '$ gives a saddle point
equation which determines the energy $E$ in terms of the asymptotics of
the solution at the initial time.
\end{enumerate}

Now, given the field asymptotics (\ref{ainneg}, \ref{ainpos}, \ref{afin}),
we can find the initial coherent state $\vert\, \{\, \ak \,\}\, \rangle$
and the most probable final coherent state, which correspond to our solution.
The saddle point conditions (\ref{bcfin}) at $t\rightarrow +\infty$ determine
the most probable final coherent state in terms of the asymptotics of the
solution (\ref{b}) found in the previous section:
\beq
b^{a }_i ({\bf k}) \: = \: g^a_i ({\bf k})\: =\: {\pi^2\sqrt{2\epsilon}\over
g}\,\sqrt{2k}\,
e^{-k}\,\left(\, e^i_a ({\bf k}) \: - \: i\,{\epsilon_{alm}k_m \over k}\,
e^i_l ({\bf k})\right) \, ,
\eeq
where the color ($a$) and polarization ($i$) indices have been restored.
The final coherent state has then the form:
\beq
\vert\,\{\, b\, ( {\bfk}\,)\, \}\,\rangle \: = \:
\exp{\left[\, \int d {\bf k} \:
b_i^a ({\bf k}) \, \hat{a}^{a \dagger}_i ({\bf k})\,\right]}
\:\vert\, 0\,\rangle \, ,
\eeq
where $\hat{a}^{a \dagger}_i ({\bf k})$ is a creation operator for a state
with polarization $i$ ($i=1,2$), and color $a$ ($a=1,2,3$)
\footnote{The operators are normalized such that
$\left[\hat{a}^a_i ({\bf p}),\hat{a}^{b \dagger}_j ({\bf k})\right]=
\delta_{ij}\,\delta^{ab}\,\delta \left({\bf p} - {\bf k}\right)$.}.
Then,
the average number of particles with momentum ${\bf k}$ in the final state is
\beq
\bar{n}^{\rm fin}_{{\bf k}} \: = \: \sum_{i,a}\: b^{a *}_i ({\bf k}) \,
b^{a}_i ({\bf k}) \: = \:
{16\,\epsilon\,\pi^4\over g^2}\, k \, e^{-2k} \, .
\eeq
The total energy of the final coherent state is
\beq
\bar{E}_{\rm fin} \: = \: \int {d {\bf k} \over (2\pi)^3}\: k\,
n^{\rm fin}_{{\bf k}} \: = \: {6\,\epsilon\,\pi^2\over g^2} \, .
\eeq
As promised, the above expression for the energy exactly coincides
with the  energy of the classical solution.
The latter is easiest calculated at $t = \phi = 0$  \cite{LUS}:
\beq
\label{energy}
E_{\rm classical} \: = \:  {12\pi \over g^2}\,\int_{-\pi /2}^{+\pi /2}dw\:
\cos^2 w \,\epsilon \: = \:
{6\,\epsilon\,\pi^2\over g^2} \, .
\eeq
This correspondence is expected.
For large $t$, the solution represents a thin spherical shell of energy
expanding with the speed of light and reduces to the superposition (\ref{afin})
of plane waves.
We find for the total average number of particles in the final state
\beq
\bar{N}_{\rm fin} \: = \:\int {d {\bf k}\over (2\pi)^3}\: n^{\rm fin}_{{\bf k}}
\: = \: {3\,\epsilon\,\pi^2\over g^2} \, .
\eeq

The saddle point conditions arising from the integration over the initial
values of the fields (\ref{bcin}) determine the initial coherent state in terms
of the asymptotics of the solution (\ref{ainneg}, \ref{f}) \cite{RST}:
\beq
\label{bcinsol}
a_i^a ({\bf k}) \: = \: f_i^a ({\bf k}) \, e^{-kT - ik\xi} \, ,
\eeq
with $f_i^a ({\bf k})$  given by (\ref{f}).

Now, the real part of $\xi$ can be removed by time translation.
The imaginary part may be fixed by requiring the average energy
of the initial state equal that of the final state \cite{RST} :
\beqa
\bar{E}_{\rm in} & = &
\int {d{\bf k}\over (2\pi)^3} k a_i^{a *} ({\bf k}) a_i^a ({\bf k})
\: = \: \int {d {\bf k}\over (2\pi)^3}\: k\, f_i^{a *} ({\bf k})
f_i^a ({\bf k}) \, e^{-2kt-2k{\rm Im} \xi} \\
& = & \bar{E}_{\rm fin} \: = \: {6 \pi^2 \epsilon \over g^2} \, .
\eeqa
Substituting (\ref{f}) for $f_i^a "$ ,
we determine the value of ${\rm Im}\,\xi $:
$$
{\rm Im}\,\xi \: = \: 1 \: - \:
\left( {45\,\pi^3\over 4}\,\epsilon \right)^{1/7} \, .
$$
We have omitted terms in this expression which are
subdominant for $\epsilon\ll 1$ .
Now the average number of particles  in the initial state is determined
to be:
$$
\bar{N}_{\rm in}\: = \:\int  {d {\bf k}\over (2\pi)^3} \:
f_i^{a *} ({\bf k}) f_i^a ({\bf k})\, e^{-2kt-2k{\rm Im} \xi} \,
\:\sim\: \epsilon^{1/7}\,\bar{N}_{\rm fin}.
$$
Our solution describes therefore a transition from a state with a
smaller number of particles, $\bar{N}_{\rm in} $,
to a state with a larger number of particles, $\bar{N}_{\rm fin} $,
their ratio being controlled by the small parameter,
\mbox{$\epsilon^{1/7} $ .}

However,
our solution does not maximize the microcanonical transition probability
(\ref{eq:microcanonical}).
 It does not give the maximum transition probability at a given energy.
The S-matrix element between our coherent states is an infinite sum of
$n$-particle scattering amplitudes:
\beqa
\label{smxsum}
\lefteqn{\langle \{ b\} \vert {\rm \hat{S}_Q}\vert \{ a \} \rangle\: \sim} \\
& &
\sum\limits_{n,m} \int \prod_{ij} d^{3}k_i\, d^{3}p_j\:
 c^* (k_1)\ldots c^* (k_n)\, d(p_1)\ldots d(p_m)\:
\langle\, k_1,\ldots k_n\,\vert\, S_Q\,\vert\, p_1,\ldots p_m\,\rangle \, .
\nonumber
\eeqa
The above calculation does not allow the determination of any
particular $n$-particle scattering amplitude entering the sum (\ref{smxsum}).
It only gives an example of a semiclassically calculable multiparticle
transition amplitude in the one-instanton sector.

\mysection{Conclusions}

In this paper,
we investigated the role of a complex time solution in Yang-Mills theory
in high-energy scattering processes.
We argued that the complex time formalism is a natural one for
describing the initial and final multiparticle states in different sectors
of the periodic vacuum \cite{RT,RSTnew}.
The free-wave asymptotics of the solution at $\vert t\vert \rightarrow\infty$
define   the initial and final coherent states, through the classical boundary
value problem  (eqs (\ref{bcin}, \ref{bcfin})) discussed in the previous
section.

In order to solve the boundary value problem however,
we considered the case of a highly-symmetric solution of the Yang-Mills
equations.
The field equations were reduced to a  quantum mechanical problem,
by exploiting the $SO(4)$-conformal symmetry of the pure gauge theory
\cite{LUS, SCH}.
In particular,
this simplification enabled us to analytically continue the \ls solution
to a complex time contour and provide some insight on the role of complex time
singularities.
The singularities were found to completely determine the topological charge
and the imaginary part of the action of the solution.
Moreover,
they turned out to obey a relation analogous to that obeyed by self-dual
Euclidean solutions.
This property is quite nontrivial on a complex time contour with complex
valued fields where the usual Bogomol'nyi bound does not apply,
and it makes the solution interesting for the problem of high energy fermion
number or chirality violation.

We found that this solution, on a suitably chosen complex time contour,
gives a saddle point contribution to a multiparticle scattering process,
for which the average number of particles in the initial and final state are
parametrically different. The ratio of the particle number in the initial and
final coherent states is controlled by a small parameter of the solution.
For reasons discussed in the Introduction,
this solution may be relevant to the problem of initial state
corrections\cite{TIN,RT,RST}.
The solution does not, however, maximize the transition probability
in the one-instanton sector at a given energy.
Thus, it can not be used to provide an upper bound on the $2\rightarrow n$
process cross-section.
It only gives an example of a semiclassically calculable multiparticle
transition amplitude with fermion number or chirality violation.

The assumption of conformal symmetry may allow a straightforward extension
of the ideas presented here to a few more complicated field equations,
coupled to the Yang-Mills equations.
Minkowski time solutions of the field equations for a scalar triplet\cite{DON}
and fermion fields \cite{DON2,MEE} coupled to gauge fields have already
appeared in the literature.
It may be interesting to investigate the properties of these solutions
on the complex time contour,
with an eye towards incorporating the additional fields of the Standard
Model in this formalism.
In particular, it may be possible to understand the process of fermion
number or chirality violation in the Dirac-Yang-Mills system
on the complex time contour.

However,
the high degree of symmetry assumed here clearly limits the scope
of the results.
The spherical symmetry ($SO(3)_{\rm rot} \subset SO(4)_{\rm conf}$)
of the solution has led us astray from the problem of high-energy $2\rightarrow
n$ processes.
The solution in Minkowski time has the form of a spherical shell of energy,
which collapses from infinity, then, at $t=0$, bounces back and expands with
the speed of light.
Clearly such a classical field configuration is a poor approximation to an
initial state of two highly energetic colliding particles.
Physical intuition would lead one to believe that a solution
with only a cylindrical symmetry might be a better candidate.

The assumption of conformal symmetry has also made less transparent
an important application of this formalism: the electroweak theory.
The mass scale $v \simeq 246\, $ GeV in the electroweak theory explicitly
breaks the classical conformal invariance of the pure gauge theory.
It is expected then that the arbitrary ``scale size'' $\rho $ of the
classical solution, set to unity in the scale-invariant analysis above,
will be fixed by the new mass scale.
In the case that the center-of-mass energy $E$ greatly exceeds the
symmetry breaking scale,
the Yang-Mills theory considered here may correctly describe the classical
behavior of the gauge sector of the electroweak theory.
Then, our results  have direct relevance to the behavior
in this energy region \cite{RSTnew}.

Despite the explicit symmetry breaking parameter $v$ in the electroweak
theory,
a conformally-symmetric solution has played a major role in the Euclidean
approach to the problem of multiparticle scattering.
As reviewed in the Introduction,
the BPST instanton forms the basis of a perturbative expansion of final
state corrections to the leading semiclassical behavior of the multiparticle
cross-section.
In fact, the BPST instanton in this case represents only the ``core''
($r\ll \rho \ll 1/M_w $) of the solution which dominates the cross-section
for energies $E\ll E_0$, a complicated approximate solution to the electroweak
gauge-Higgs field equations \cite{AFF}.
Corrections to the core behavior at large distances lead to higher order
corrections in $E/E_0$ in the leading semiclassical behavior
(\ref{eq:holygrail}) of the total cross-section \cite{KRT,MAT}.

The complex time solution presented here may also be considered
the ``core'' of a constrained solution in a spontaneously-broken
gauge theory.
At distances larger than $1/v$,
the solution would have the exponentially-damped behavior characteristic
of a massive gauge field.
The complex time approach here differs however in at least one important
respect.
In the Euclidean approach,
there are {\it no} exact finite-action solutions to the electroweak gauge-Higgs
field equations,
as may be established by a simple scaling argument.
In this case,
the constrained expansion is a device to obtain the {\it approximate} solutions
which provide the dominant semiclassical contribution to scattering
amplitudes\cite{AFF}.
In the present approach however,
nothing prevents the existence of an {\it exact} solution to the Minkowski
gauge-Higgs equations.
Such a solution would represent an additional saddle point contribution
to a transition amplitude.
The effect of symmetry-breaking on the solution presented in this paper
has yet to be explored.


\newpage

\begin{thebibliography}{99}
\bibitem{JAC}R. Jackiw and C. Rebbi, {\em Phys. Rev. Lett.}
{\bf 37} (1976) 172.
\bibitem{CDG}C.G. Callan, R. Dashen and D.J. Gross,
{\em Phys. Lett.} {\bf 63B} (1976) 334.
\bibitem{THO}G. 'tHooft, {\em Phys. Rev.} {\bf D14} (1976) 3432;
{\em Phys. Lett.} {\bf 37B} (1976) 8.
\bibitem{RIN}A. Ringwald, {\em Nucl. Phys.} {\bf B330} (1990) 1.
\bibitem{ESP}O. Espinosa, {\em Nucl. Phys.} {\bf B343} (1990) 310.
\bibitem{early}For an early review, see M.P. Mattis and E. Mottola, (eds.),{\em
Baryon Number Violation at the SSC? Proceedings}, Workshop, Santa Fe, USA,
April 27-30, 1990 (Singapore: World Scientific, 1990)
\bibitem{MAT}For a recent review, see M.P. Mattis, {\em Phys. Rep.} {\bf 214}
(1992) 159.
\bibitem{BPST}A.A. Belavin, A.M. Polyakov, A.S. Schwarz,  and Yu.S. Tyupkin,
{\em Phys. Lett.} {\bf 59B} (1975) 85.
\bibitem{KRT}S. Yu. Khlebnikov, V.A. Rubakov and P. Tinyakov,
{\em Nucl. Phys.} {\bf B350} (1991) 441; \\
{\em Mod. Phys Lett.} {\bf A5} (1990) 1983.
\bibitem{ZAK}V.I. Zakharov, Nucl. Phys. {\bf B353} (1991) 683;
{\em ibid.} {\bf B377} (1992) 501.
\bibitem{MAG}M. Maggiore and M. Shifman, {\em Nucl. Phys.} {\bf B371} (1992)
177; \\
{\em Phys. Rev.} {\bf D46} (1992) 3550.
\bibitem{GOU}T.M. Gould, {\em Phys.Lett.} {\bf 282B} (1992) 149.
\bibitem{KT}S.Yu. Khlebnikov and P.G. Tinyakov, {\em Phys.Lett} {\bf 269B}
       (1991) 149;\\
       P.B. Arnold and M.P. Mattis, {\em Phys.Rev.} {\bf D44} (1991) 3650.
\bibitem{MUE}A. Mueller, {\em Nucl.Phys.} {\bf B348} (1991) 310,
{\em ibid.} {\bf B353} (1991) 44.
\bibitem{MMY}M.P. Mattis, L. McLerran, and L.G. Yaffe,
{\em Phys. Rev.} {\bf D45} (1992) 4294.
\bibitem{TIN}P.G. Tinyakov, {\em Phys. Lett.} {\bf B284} (1992) 410.
\bibitem{RST}V.A. Rubakov, D.T. Son and P.G. Tinyakov,
  {\em Phys.Lett} {\bf 287B} (1992) 342.
\bibitem{RT}V.A. Rubakov and P.G. Tinyakov, {\em Phys. Lett.} {\bf B279} (1992)
165.
\bibitem{RSTnew}V.A. Rubakov, D.T. Son and P.G. Tinyakov,
{\em An Example of Semiclassical Instanton-like Scattering: 1+1 Dimensional
Sigma Model}, Minnesota preprint, TPI-MINN-92-66-T, hep-ph 9212309,
December 1992.
\bibitem{KYA}A.B. Kyatkin, {\em Example of  Instanton-Like Scattering
in $\lambda \, \phi^4$-Theory}, Johns Hopkins preprint, to appear.
\bibitem{KRT3}S.Yu. Khlebnikov, V.A. Rubakov, P.G. Tinyakov,
{\em Nucl. Phys.} {\bf B367} (1991) 334.
\bibitem{GPY}D.J. Gross, R.D. Pisarski, and L.G. Yaffe,
{\em Rev. Mod. Phys.} {\bf 53} (1981) 43.
\bibitem{HSU}S.D.H. Hsu, {\em Phys. Lett.} {\bf 261B} (1991) 81.
\bibitem{FKS}E. Farhi, V.V. Khoze and R. Singleton,Jr., {\em Minkowski Space
Non-Abelian Classical Solutions with Non-Integer Topological
Number}, MIT preprint, CTP 2138, December 1992.
\bibitem{LUS}M. L\"uscher, {\em Phys.Lett} {\bf 70B} (1977) 321.
\bibitem{SCH}B. Schechter, {\em Phys.Rev.} {\bf D16} (1977) 3015.
\bibitem{GR}I.S. Gradstein and I.M. Ryzhik, {\em Table of Integrals, Series and
Products},Academic Press, N.-Y., 1980.
\bibitem{JNR}R. Jackiw, C. Nohl, and C. Rebbi, {\em Phys. Rev.} {\bf D15}
(1977) 1642.
\bibitem{WIT}E. Witten, {\em Phys. Lett.} {\bf 117B} (1982) 324.
\bibitem{BOG}E.B. Bogomol'nyi, {\em Sov. J. Nucl. Phys.} {\bf 24} (1976) 449.
\bibitem{LUS2}M. L\"uscher, {\em Nucl. Phys.} {\bf B140} (1978) 429.
\bibitem{DON}J. Doneux, Y. St.-Aubin, and L. Vinet, {\em Phys. Rev. }
{\bf D24}(1981) 3179.
\bibitem{DON2}J. Doneux, Y. St.-Aubin, and L. Vinet, {\em Phys. Rev. }
{\bf D25}(1982) 484.
\bibitem{MEE}K. Meetz, {\em Z. Phys. }{\bf C6} (1980) 41.
\bibitem{AFF}I. Affleck, {\em Nucl. Phys.} {\bf B191} (1981) 445.

\end{thebibliography}
\end{document}